\documentclass[aps,prb,twocolumn,groupaddress,showpacs]{revtex4}
\usepackage{amssymb}
\usepackage{graphicx}
\usepackage{amsmath}
\usepackage{subfigure}

\begin{document}

\preprint{}
\title{Charge Density Wave Instability and Soft Phonon in $A$Pt$_3$P ($A$=Ca, Sr, and La)}

\author{Hui Chen}

\author{Xiaofeng Xu}

\author{Chao Cao}
  \email[E-mail address: ]{ccao@hznu.edu.cn}

\author{Jianhui Dai}
  \email[E-mail address: ]{daijh@hznu.edu.cn}
  \affiliation{Condensed Matter Physics Group,
  Department of Physics, Hangzhou Normal University, Hangzhou 310036, China}

\date{Jul. 16, 2012}

\begin{abstract}
The electronic and phonon properties of the platinum pnictide
superconductors $A$Pt$_3$P ($A$=Ca, Sr, and La) were studied using
first-principles calculations. The spin-orbit coupling effect is
significant in LaPt$_3$P but negligible in CaPt$_3$P and SrPt$_3$P.
Moreover, SrPt$_3$P has been demonstrated to exhibit an unexpected
weak charge-density-wave (CDW) instability which is neither simply
related to the Fermi-surface nesting nor to the momentum-dependent
electron-phonon coupling alone.
The instability is absent in CaPt$_3$P and can be quickly suppressed
by the external pressure, accompanied with decreases in the
phonon softening and BCS $T_c$. Our results suggest SrPt$_3$P as a
rare example where superconductivity is enhanced by the CDW
fluctuations.
\end{abstract}


\maketitle 

\section{Introduction}
The discovery of iron-based high-Tc superconductors has stimulated
considerable interests in the condensed matter physics society.
Sharing the similar layered square lattice crystal structures, both
iron-pnictides/chalcogenides and cuprates exhibit the long-range
magnetic order in parent compounds and the unconventional
superconductivity (SC) upon doping
\cite{pcdai_nature_453_899,dong_epl_83_27006,PhysRevLett.101.257003,
PhysRevLett.102.247001,PhysRevLett.102.177003,arXiv:1102.0830}. In
these $3d$ transition-metal based materials, the electron
correlation effect is considered to play a role in the formation of
the spin-density-wave (SDW) instability while the electron pairing
may be mediated or enhanced by the spin fluctuations.

\begin{figure}
  \subfigure[Crystal Structure]{
    \includegraphics[scale=0.12]{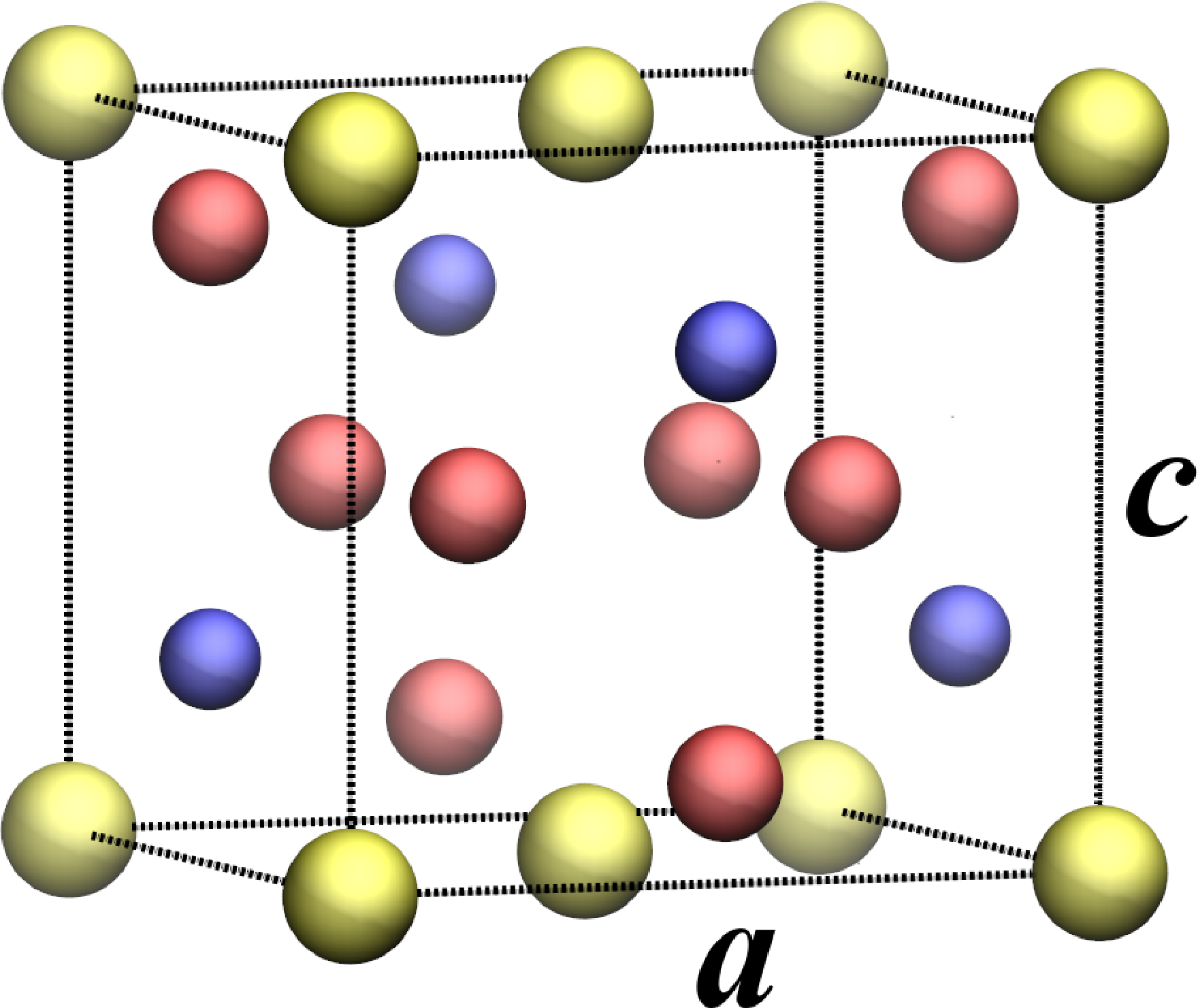}}
  \subfigure[Pt$_2$P layer]{
    \includegraphics[scale=0.12]{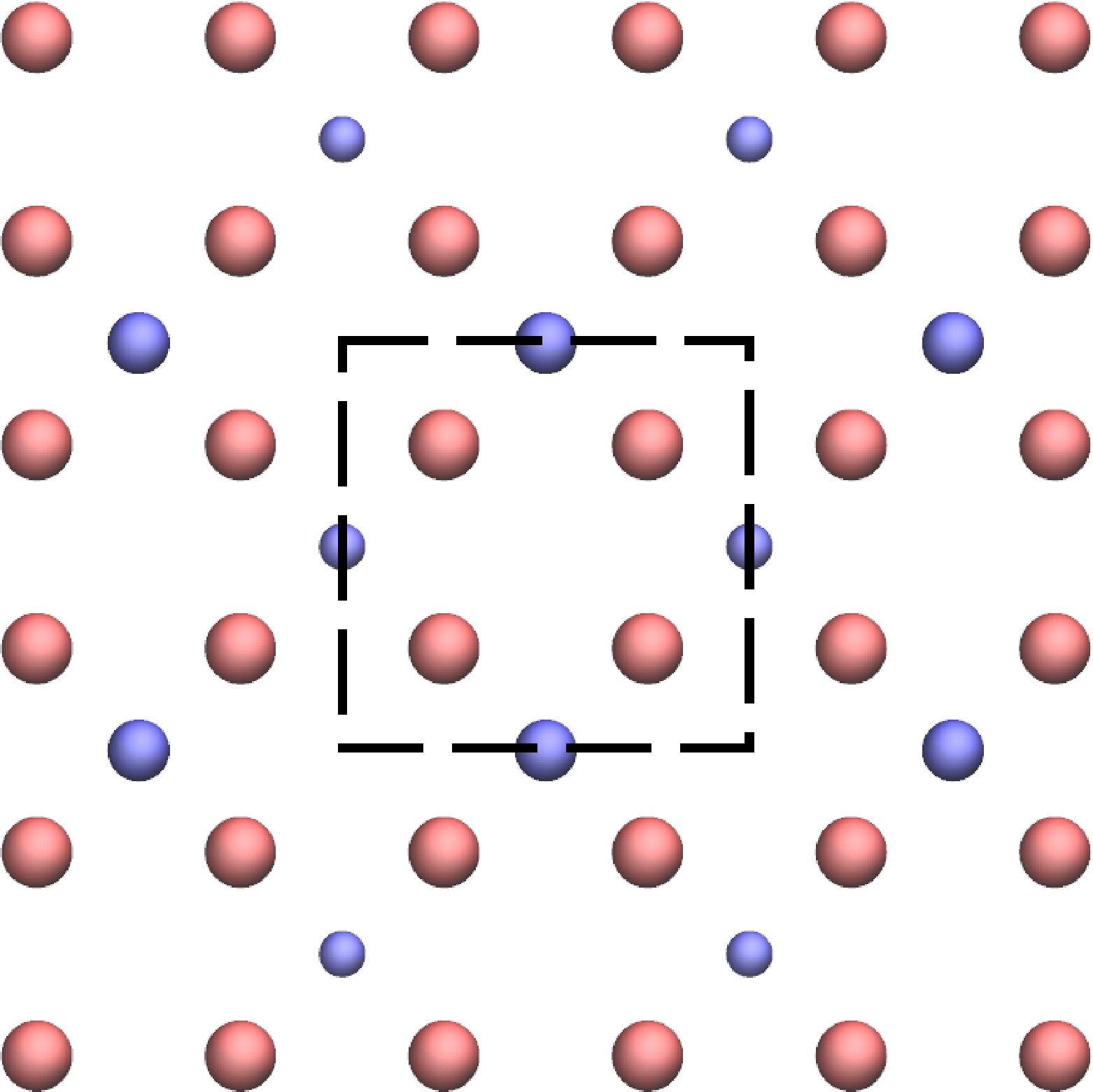}}
  \caption{(color online) Crystal structure of $A$Pt$_3$P compounds. (a) Crystal structure of $A$Pt$_3$P. Yellow atoms are $A$ (Ca, Sr, or La) atoms; red atoms are platinum; and blue atoms are phosphorus. (b) Pt$^{\mathrm{II}}$ and P atoms form a Pt$_2$P layer that is geometrically similar to the iron-arsenic layer in the iron-pnictides. Larger blue atoms denote the P atoms above the Pt plane and smaller blue atoms are the ones below the Pt plane.\label{fig_geometry}}
\end{figure}

Recently, several families of platinum-pnictide based
superconductors(SrPtAs\cite{JPSJ.80.055002},
SrPt$_2$As$_2$\cite{JPSJ.79.123710},
$A$Pt$_3$P\cite{arXiv:1205.1589}) have been reported. Among these
compounds, $A$Pt$_3$P ($A$=Ca, Sr, and La) have shown considerable
peculiarity. The crystal structures of these $5d$ electron-based
pnictides are experimentally determined to be anti-pevroskite with
$\sqrt{2}\times\sqrt{2}\times1$ distortion (FIG.
\ref{fig_geometry}). Due to the distortion, the platinum atoms take
two different sites Pt$^{\mathrm{I}}$ and Pt$^{\mathrm{II}}$, while
the Pt$^{\mathrm{II}}$ and P atoms form a Pt$_2$P layer that
resembles the iron-pnictide layers in the iron-based
superconductors(FIG. \ref{fig_geometry}(b)). A particularly
interesting feature of this family is the significant variation of
$T_c$, i.e., $T_c=1.5$K, $6.6$K, and $8.4$K for $A=$La, Ca, and
Sr, respectively. Among them, SrPt$_3$P seems to be very special, as
it has not only the highest $T_c$ in the $5d$-electron based
superconductors, but also a large $2\Delta_0/k_BT_c\sim 5$,
manifesting strong-coupling BCS superconductivity. It also exhibits
a substantially lower electron specific heat coefficient $\gamma$
than its homologue CaPt$_3$P and a notable anomaly near 10K slightly
above $T_c$.

There are two major distinctions between the $3d$ and $5d$
transition metal pnictides. First, the $5d$-orbitals are usually
more extended so that the Coulomb interaction is relatively weak.
Second, the spin-orbit coupling (SOC) in $5d$-electrons is much
stronger, as exemplified by the chemically similar compound
SrPtAs\cite{JPSJ.80.055002}. In order to understand the puzzling
features observed in the $A$Sr$_3$P family, it is necessary and
informative to clarify the spin-orbit coupling (SOC) and the
electron-phonon coupling effects in the electronic structure of
these materials.

In this article, we present our latest first principles results of
$A$Pt$_3$P ($A$=Ca, Sr, and La). We show that while spin-orbit 
coupling effect is negligible in SrPt$_3$P and CaPt$_3$P, it 
significantly affects the electronic structure of LaPt$_3$P around 
the Fermi level $E_F$. An unexpected weak charge-density-wave (CDW)
instability develops in the SrPt$_3$P which is neither simply
related to the Fermi-surface nesting nor to the momentum-dependent
electron-phonon coupling alone. The instability is absent in 
CaPt$_3$P and can be quickly suppressed by the external pressure, 
accompanied with decreases in the BCS $T_c$. The CDW also reduces
the electron density of states (DOS) around $E_F$ for SrPt$_3$P, 
leading to the anomalous specific heat behavior around 10K.

The rest of this paper is organized as follows. In the next section,
we briefly introduces the calculation method and details, followed
by a section where we present in detail the electronic structure of
$A$Pt$_3$P ($A$=Sr, Ca, and La), the phonon properties of SrPt$_3$P
compared with CaPt$_3$P, as well as electron-phonon interactions 
and theoretical $T_c$ of SrPt$_3$P. We finally summarize and draw 
conclusions in section III. 

\section{Method and Calculation Details}
To study the electronic structure of $A$Pt$_3$P, we employed the plane
-wave basis projected augmented wave density functional (DFT) method 
implemented in the Vienna ab-initio simulation package (VASP)
\cite{VASP_1,VASP_2}. The phonon properties were obtained using the 
frozen phonon method implemented in PHONOPY\cite{phonopy}. To ensure the
calculation convergency to 1meV/cell, a 480 eV energy cutoff to the
plane-wave basis and a $8\times8\times8$ Monkhorst-Pack $\mathbf{k}$-grid 
were employed to perform structural relaxation until the maximum force on
individual atoms is smaller than 0.001 eV/\AA ~and the internal
stress smaller than 0.01 kbar. The DOS results are obtained using
$16\times16\times16$ $\Gamma$-centered $\mathbf{k}$-grid and the tetrahedra
method. For electron-phonon coupling calculations and superconducting 
$T_c$ estimation, we used density functional perturbation theory 
implemented in the \emph{Quantum Espresso} package\cite{pwscf_1}, where
the electron-phonon coupling constants were evaluated with $4\times4\times4$
$\Gamma$-centered phonon $\mathbf{q}$-grid and $32\times32\times32$ $\Gamma$-
centered $\mathbf{k}$-mesh.

\section{Results and Discussion}
\subsection{Geometry and Electronic Structures}

\begin{table}
 \begin{tabular}{c|c|c|c}
  \hline
          &  SrPt$_3$P & CaPt$_3$P & LaPt$_3$P \\
  \hline
 $a$ (\AA)           &  5.8788 (5.8852) & 5.7264 (5.7284) & 5.8187 (5.8221) \\
 $c$ (\AA)           &  5.4313 (5.4339) & 5.4507 (5.4618) & 5.5230 (5.5273) \\
 $z^{\mathrm{I}}_{\mathrm{Pt}}$ &  0.1376 (0.1394) & 0.1370 (0.1382) & 0.1435 (0.1443) \\
 $z_{\mathrm{P}}$    &  0.2767 (0.2754) & 0.2686 (0.2672) & 0.2717 (0.2707) \\
  \hline
 \end{tabular}
 \caption{Optimized geometry of $A$Pt$_3$P.
 $z^{\mathrm{I}}_{\mathrm{Pt}}$ denotes the fractional $z$-coordinate of Pt$^{\mathrm{I}}$
site. The numbers outside (inside) the brackets are obtained without
(with) SOC. \label{tab_geometry}}
\end{table}

\begin{figure*}[htp]
  \subfigure[] {
    \rotatebox{270}{\scalebox{0.43}{\includegraphics{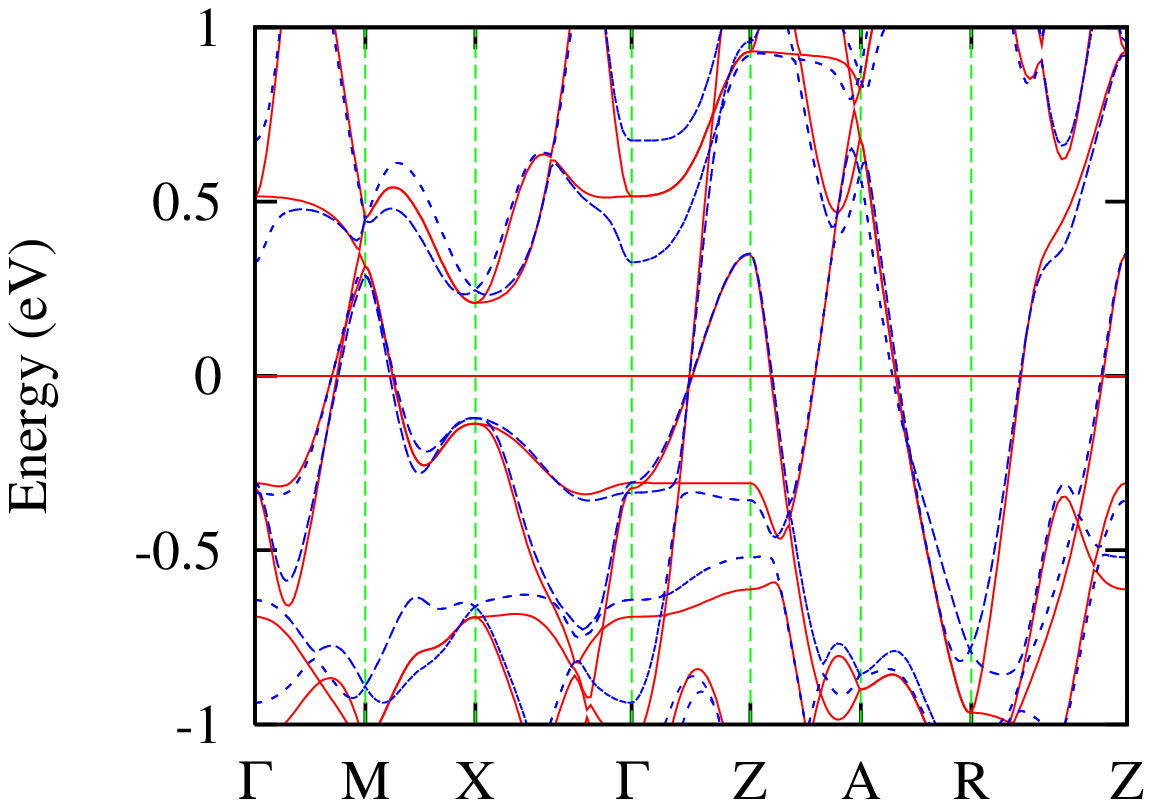}}}}
  \subfigure[] {
    \rotatebox{270}{\scalebox{0.43}{\includegraphics{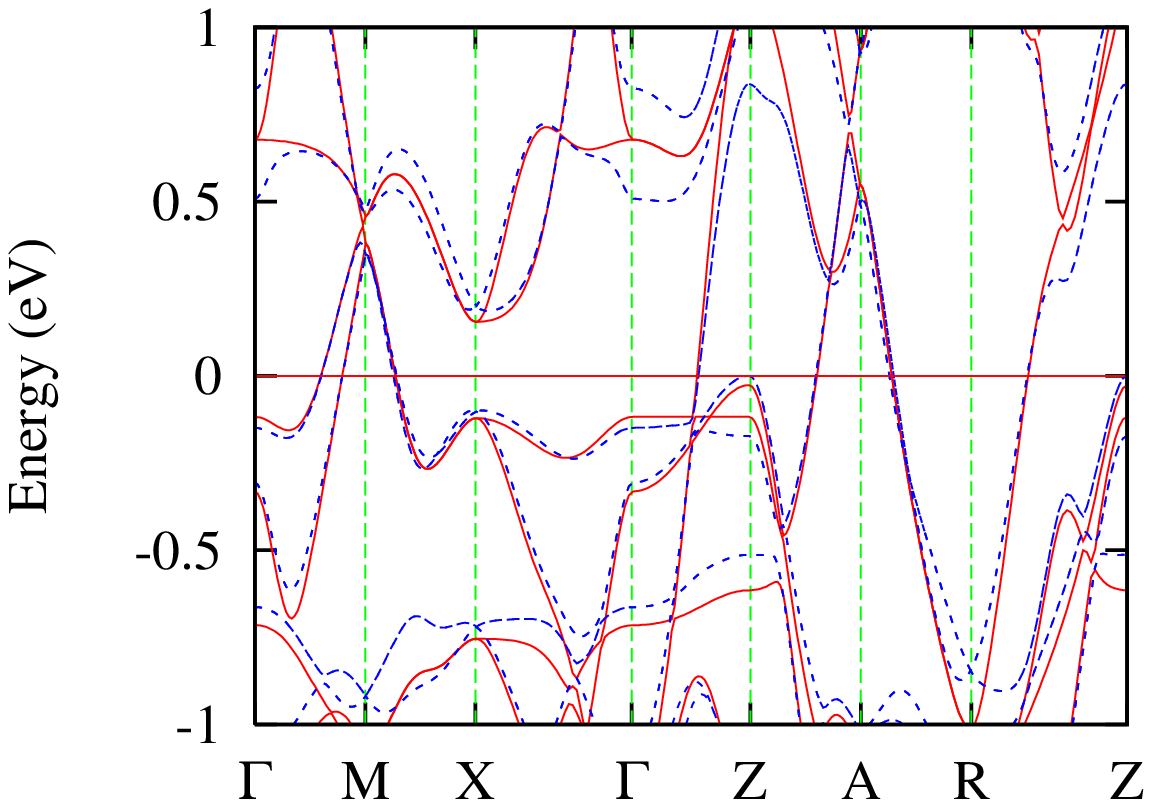}}}}
  \subfigure[] {
    \rotatebox{270}{\scalebox{0.43}{\includegraphics{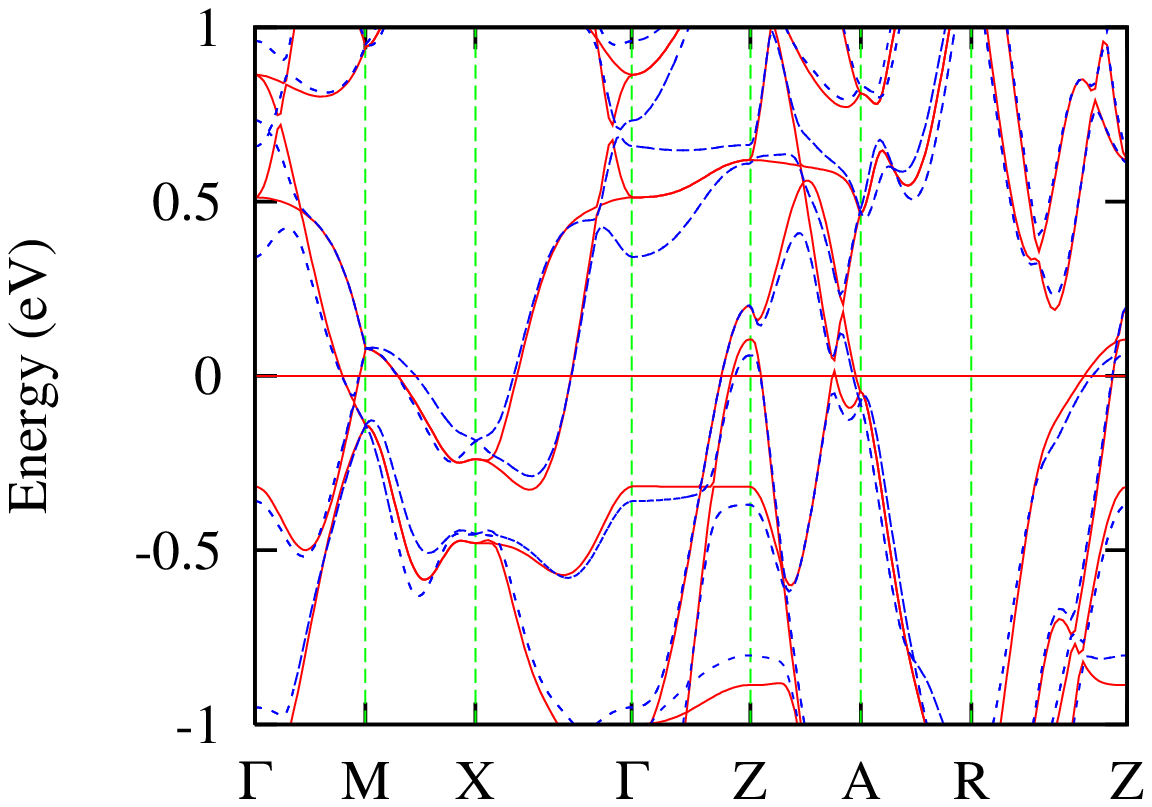}}}}
  \subfigure[SrPt$_3$P] {
    \rotatebox{270}{\scalebox{0.43}{\includegraphics{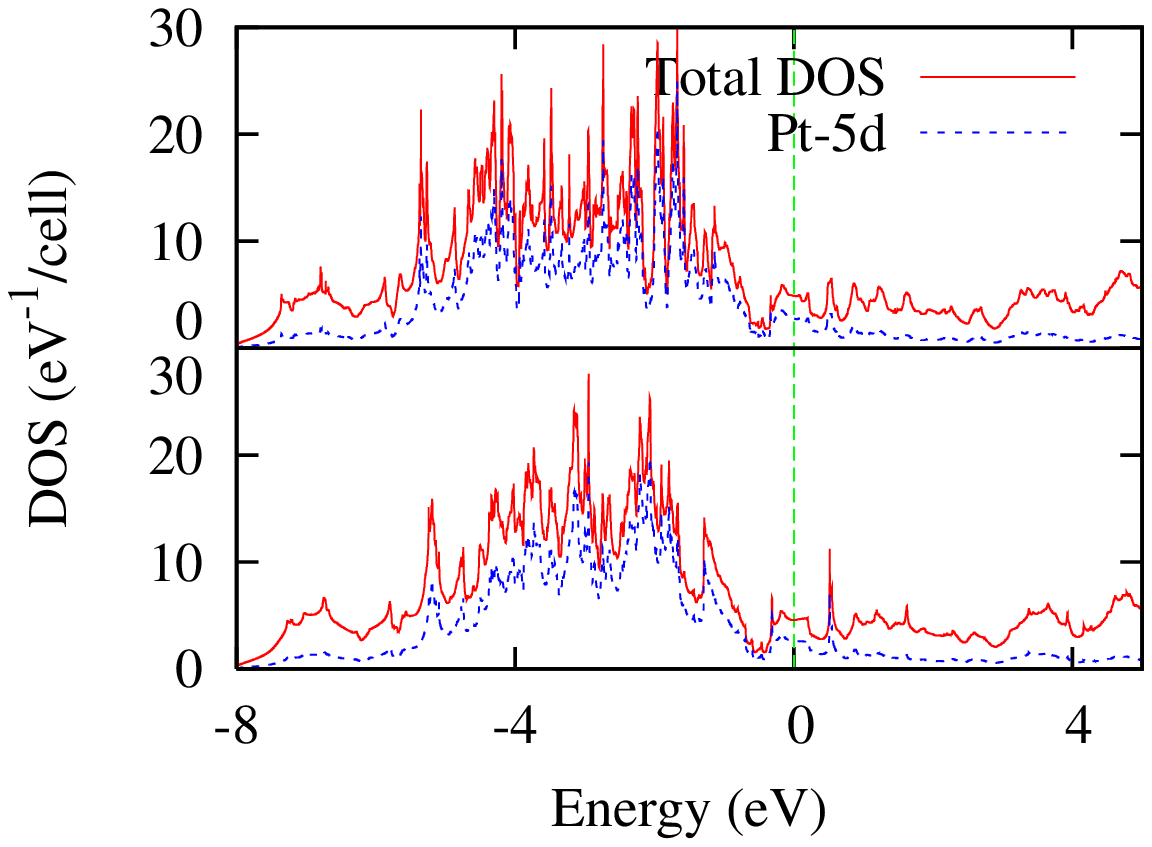}}}}
  \subfigure[CaPt$_3$P] {
    \rotatebox{270}{\scalebox{0.43}{\includegraphics{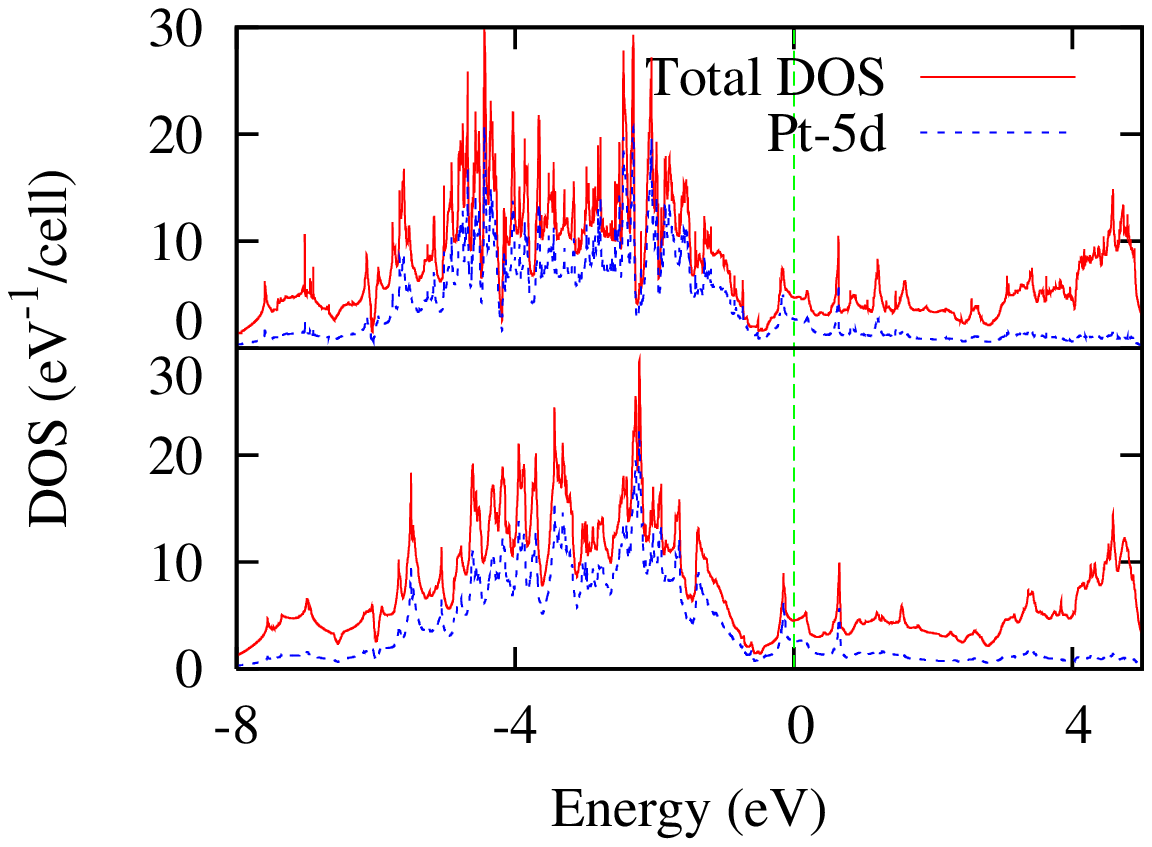}}}}
  \subfigure[LaPt$_3$P] {
    \rotatebox{270}{\scalebox{0.43}{\includegraphics{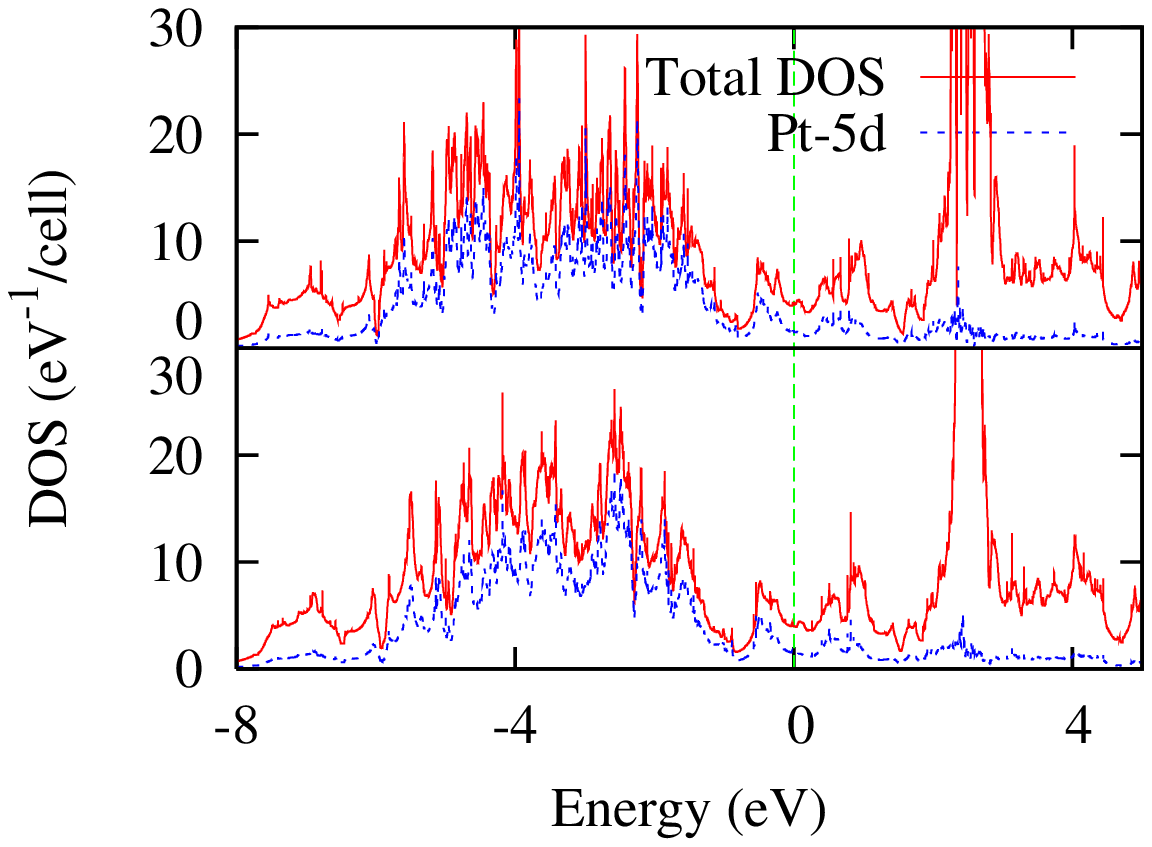}}}}
  \caption{Band structure and DOS of (a)(d) SrPt$_3$P, (b)(e) CaPt$_3$P, and (c)(f) LaPt$_3$P at ambient pressure (0 GPa) with and without SOC. (a)-(c) are band structures, where red solid lines and blue dashed lines are obtained without and with SOC, respectively.  (d)-(f) show DOS and its projection onto Pt-5d orbitals, with upper panels and lower panels showing the results with and without SOC, respectively.\label{fig_bsdos}}
\end{figure*}

\begin{figure}[htp]
  \subfigure[Sr Sheet 1]{
    \includegraphics[width=2.5cm]{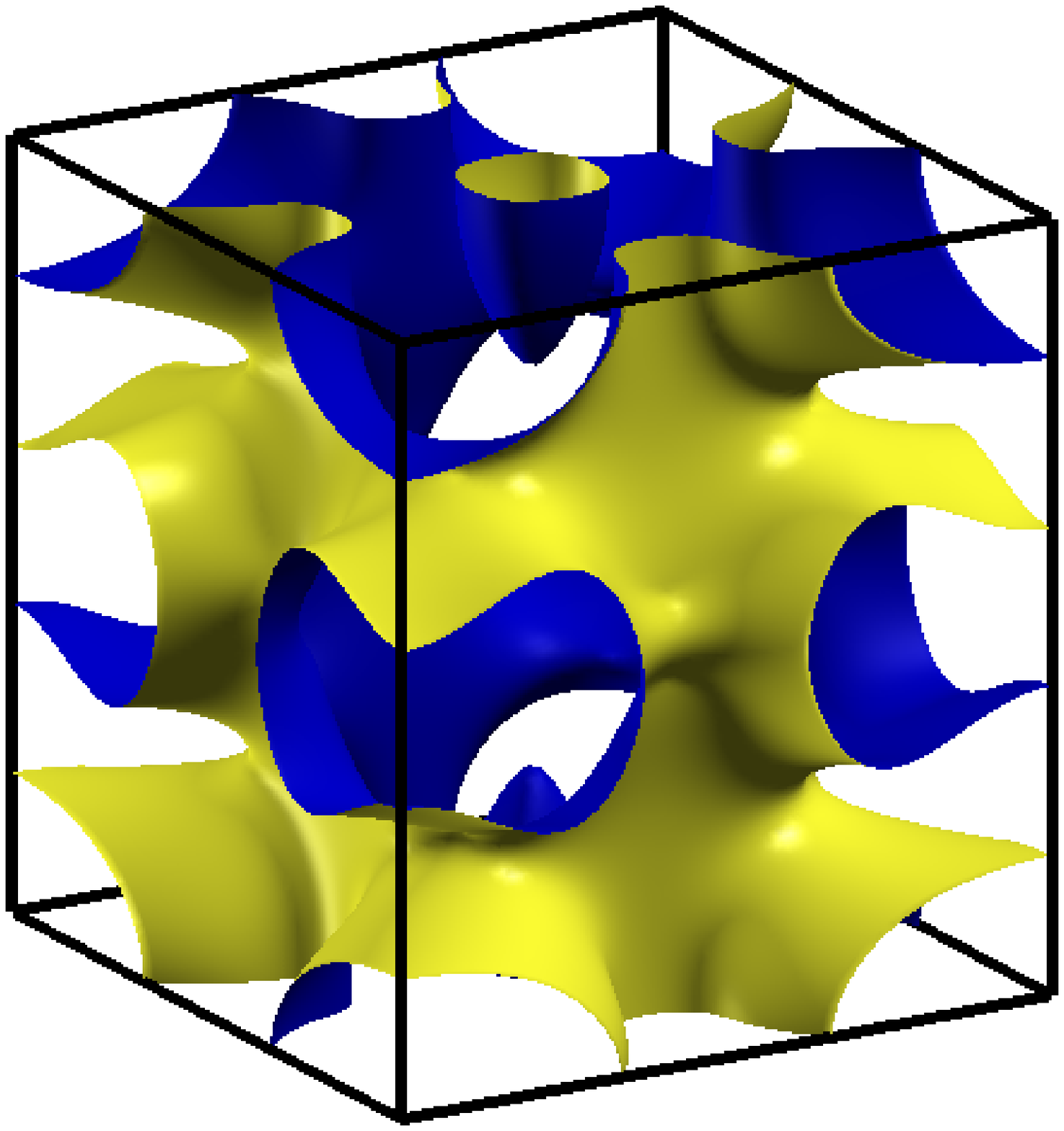}}
  \subfigure[Sr Sheet 2]{
    \includegraphics[width=2.5cm]{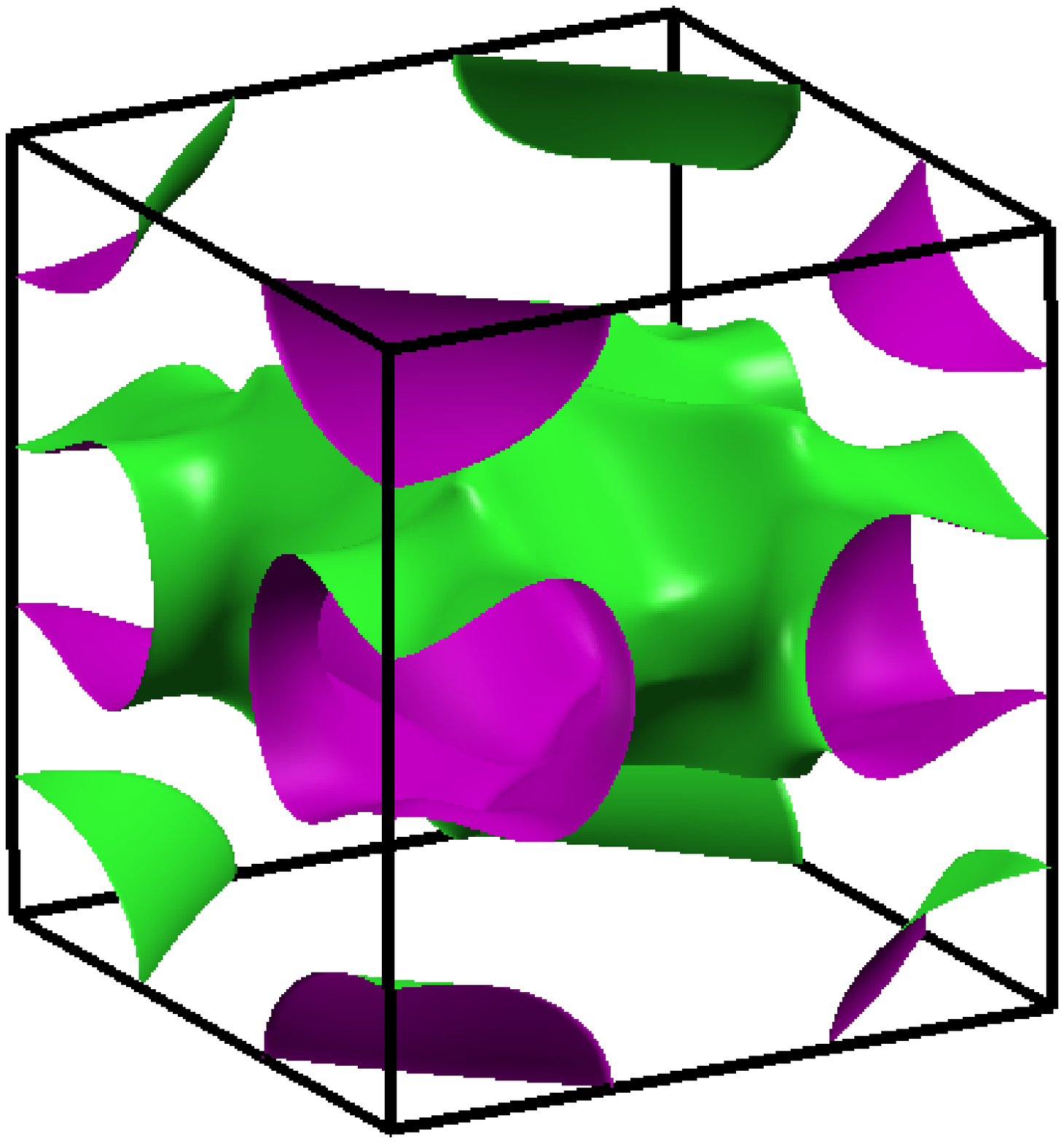}}
  \subfigure[SrPt$_3$P] {
    \includegraphics[width=3cm]{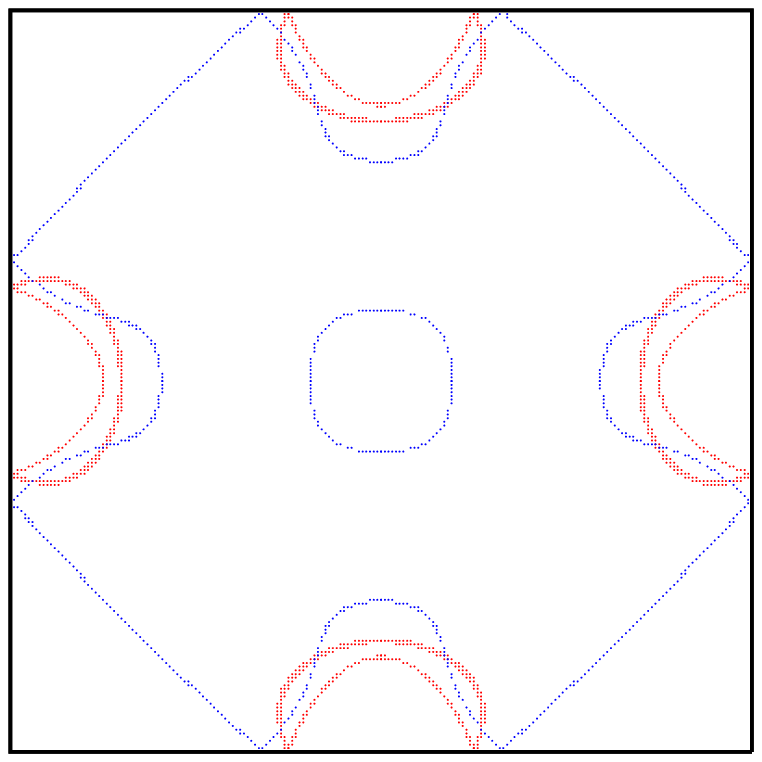}}
  \subfigure[Ca Sheet 1]{
    \includegraphics[width=2.5cm]{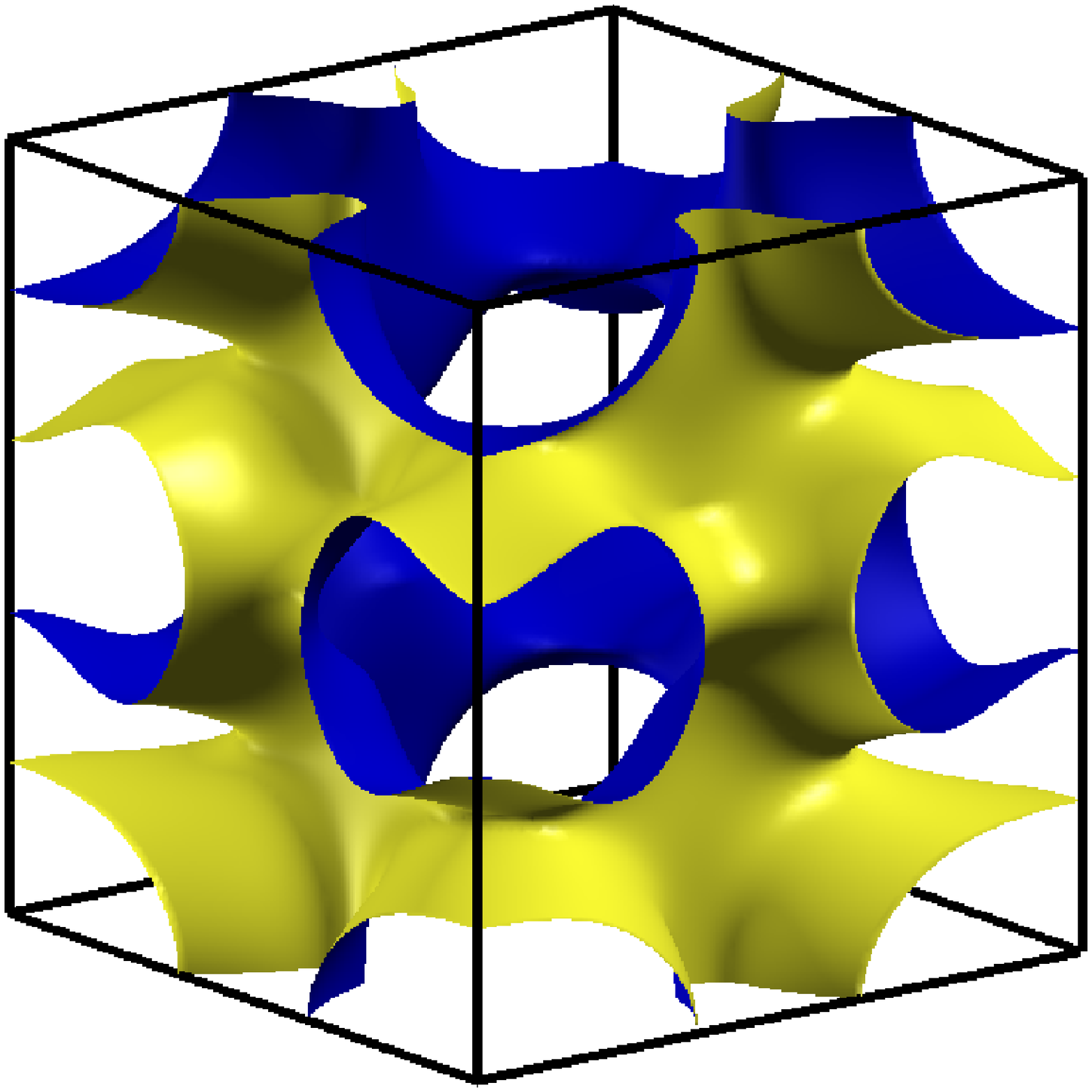}}
  \subfigure[Ca Sheet 2]{
    \includegraphics[width=2.5cm]{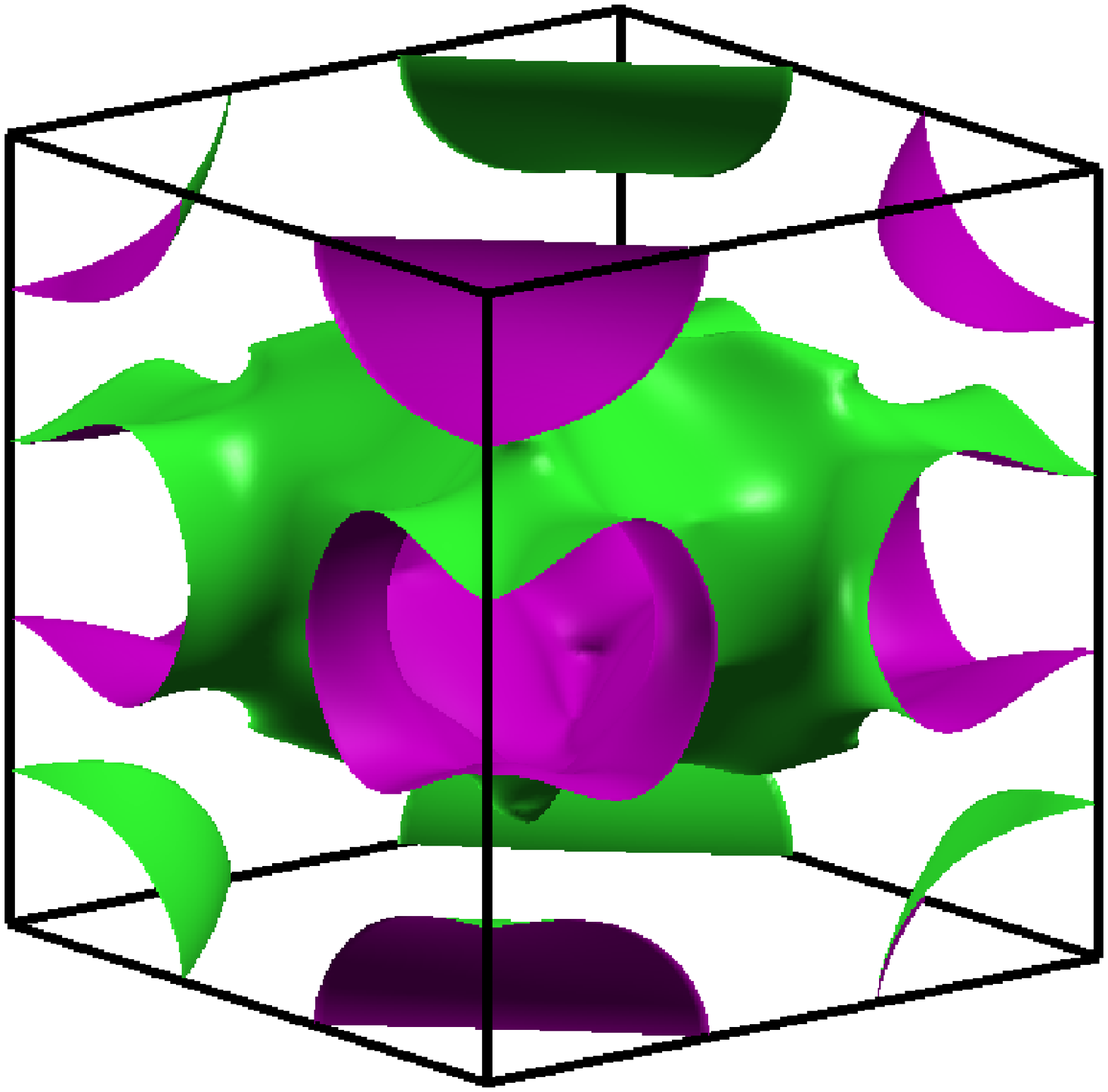}}
  \subfigure[CaPt$_3$P] {
    \includegraphics[width=3cm]{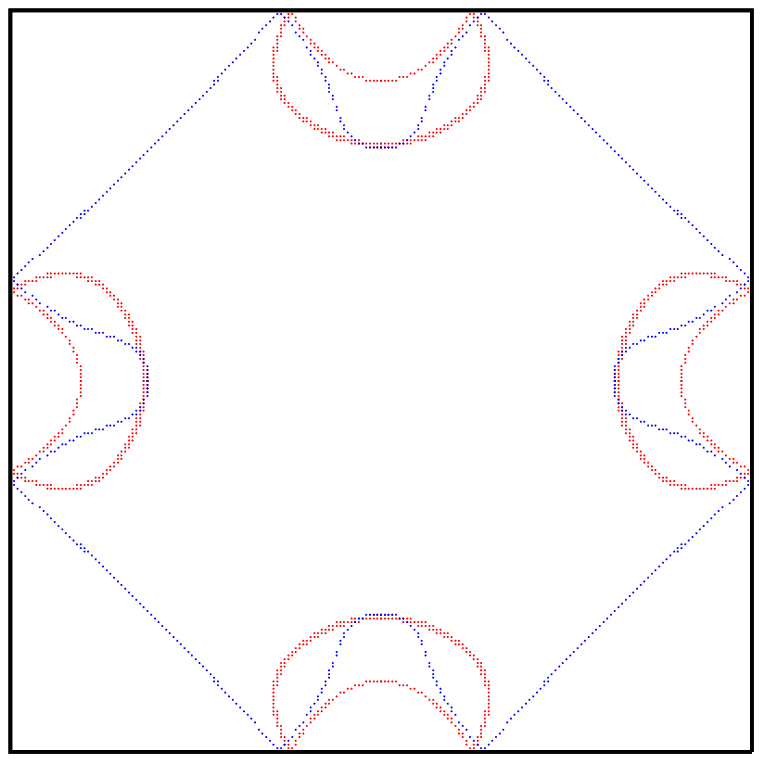}}
  \caption{(color online)Fermi surface sheets of (a-b) SrPt$_3$P and (d-e) CaPt$_3$P at 0 GPa. Fermi surface cross section of (e) SrPt$_3$P and (f) CaPt$_3$P at $k_z=0.0$ (red) and $k_z=0.5$ (blue). The centers of the squares are $\Gamma$(Z) and the corners are $X$. Notice that CaPt$_3$P has one less fermi surface sheet around $Z$. \label{fig_fs}}
\end{figure}

\begin{figure}[htp]
  \rotatebox{270}{\scalebox{0.55}{\includegraphics{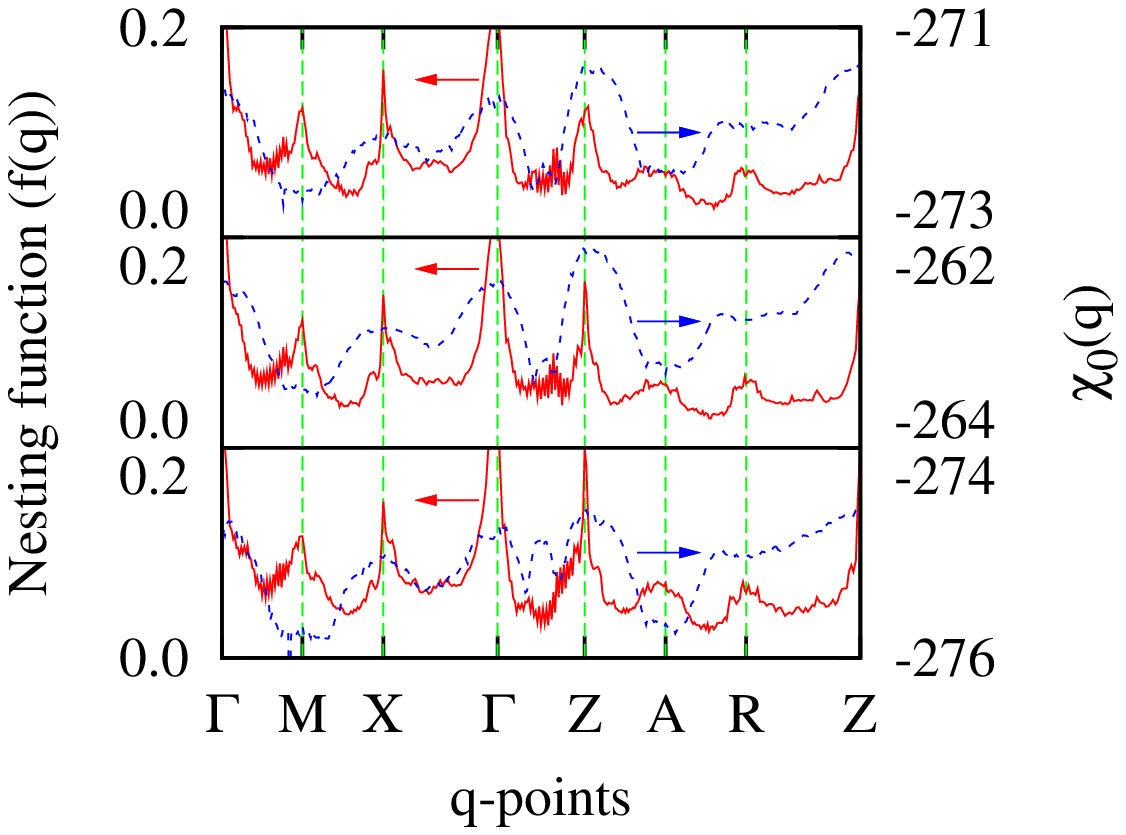}}}
  \caption{(color online) Fermi surface nesting function ($f(\mathbf{q})$, red solid lines) and electron response function ($\chi_0(\mathbf{q})$, blue dashed lines) for SrPt$_3$P at 0GPa (top panel) and 5GPa (middle panel), as well as CaPt$_3$P at 0GPa (bottom panel).\label{fig_fsnest}}
\end{figure}

The optimized structural parameters are summarized in TABLE
\ref{tab_geometry}, which agree very well with the experimental
results\cite{arXiv:1205.1589} ($<$3\% difference). Thus the
application of the DFT(+SOC) method is guaranteed in these materials
with a relatively weak Coulomb interaction. We then explore the
electronic structures shown in FIG. \ref{fig_bsdos}. For both
SrPt$_3$P and CaPt$_3$P, the SOC does not affect the electronic
structure around the Fermi energy $E_F$, as no band splitting due to
the SOC is observed within $\sim E_F\pm$500 meV range. Hence the
Fermi surfaces and consequently the SC of these two
compounds are not affected by the SOC effect. The situation is,
however, quite different in the case of LaPt$_3$P, as it exhibits a
significant band splitting due to the SOC effect from $M$ to $X$
around $E_F$, where an extra Fermi surface sheet emerges. This
observation accounts for the significant reduction of $T_c$ in
LaPt$_3$P because the SOC-induced band splitting breaks the spin
symmetry and thus reduces the pairing strength of the electrons.
Consequently, LaPt$_3$P may be a new member of the unconventional
superconductors like SrPtAs\cite{arXiv:1111.5058}
which deserves further experimental study. In the rest of our
discussions, we will mainly focus on SrPt$_3$P and CaPt$_3$P.

Similarities between their electronic structures are expected, since
both Sr and Ca are iso-valent alkaline earth elements with two
outmost s-electrons. Projected density of states (PDOS) plots (FIG. \ref{fig_bsdos}(d)-(e))
suggest that the DOS around $E_F$ in both compounds are dominated by
Pt-$5d$ and P-$2p$ orbitals, while the contribution from Sr or Ca
orbitals are minimum. However, the band structures of SrPt$_3$P and
CaPt$_3$P show two significant differences around $E_F$. First, one
of the hole bands crossing $E_F$ around the $Z$ point in SrPt$_3$P
is missing in CaPt$_3$P. Thus CaPt$_3$P has one less Fermi surface
sheets than SrPt$_3$P(FIG. \ref{fig_fs}). Second, two bands cross
$E_F$ at approximately the same $\mathbf{k}$ point in SrPt$_3$P from
$\Gamma$ to $M$, while these crossing $\mathbf{k}$ points are well
separated in CaPt$_3$P.

\begin{figure}[htp]
 \centering
  \begin{minipage}[t]{4.5 cm}
  \subfigure[SrPt$_3$P 0GPa] {
    \rotatebox{270}{\scalebox{0.35}{\includegraphics{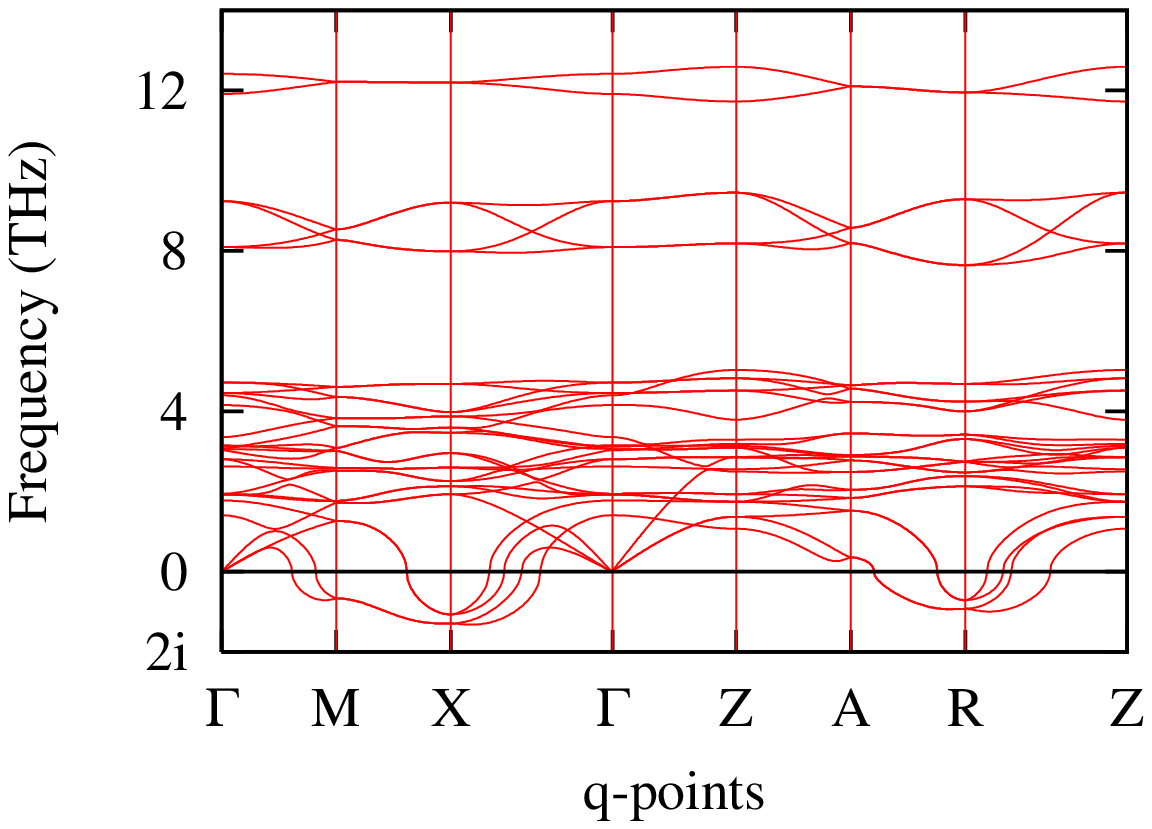}}}}
  \subfigure[SrPt$_3$P 3GPa] {
    \rotatebox{270}{\scalebox{0.35}{\includegraphics{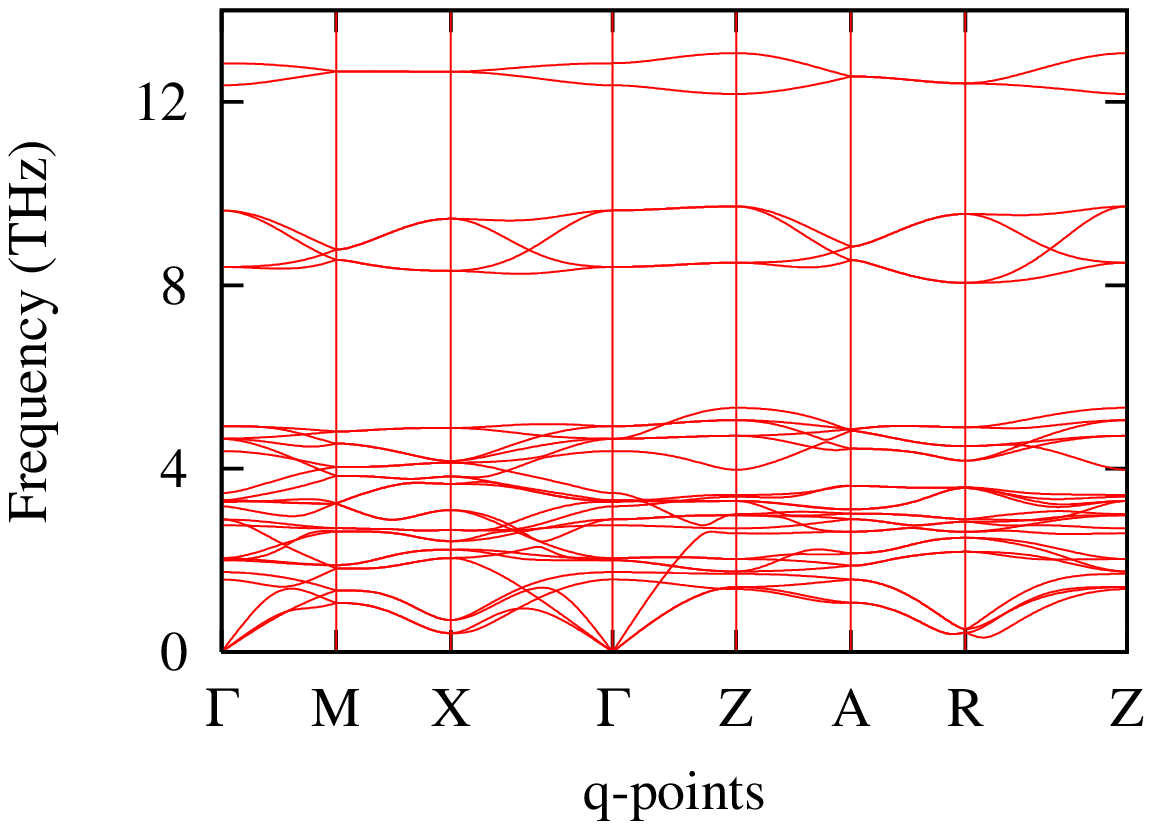}}}}
  \end{minipage}
  \begin{minipage}[t]{4 cm}
  \subfigure[Phonon DOS] {
    \rotatebox{270}{\scalebox{0.37}{\includegraphics{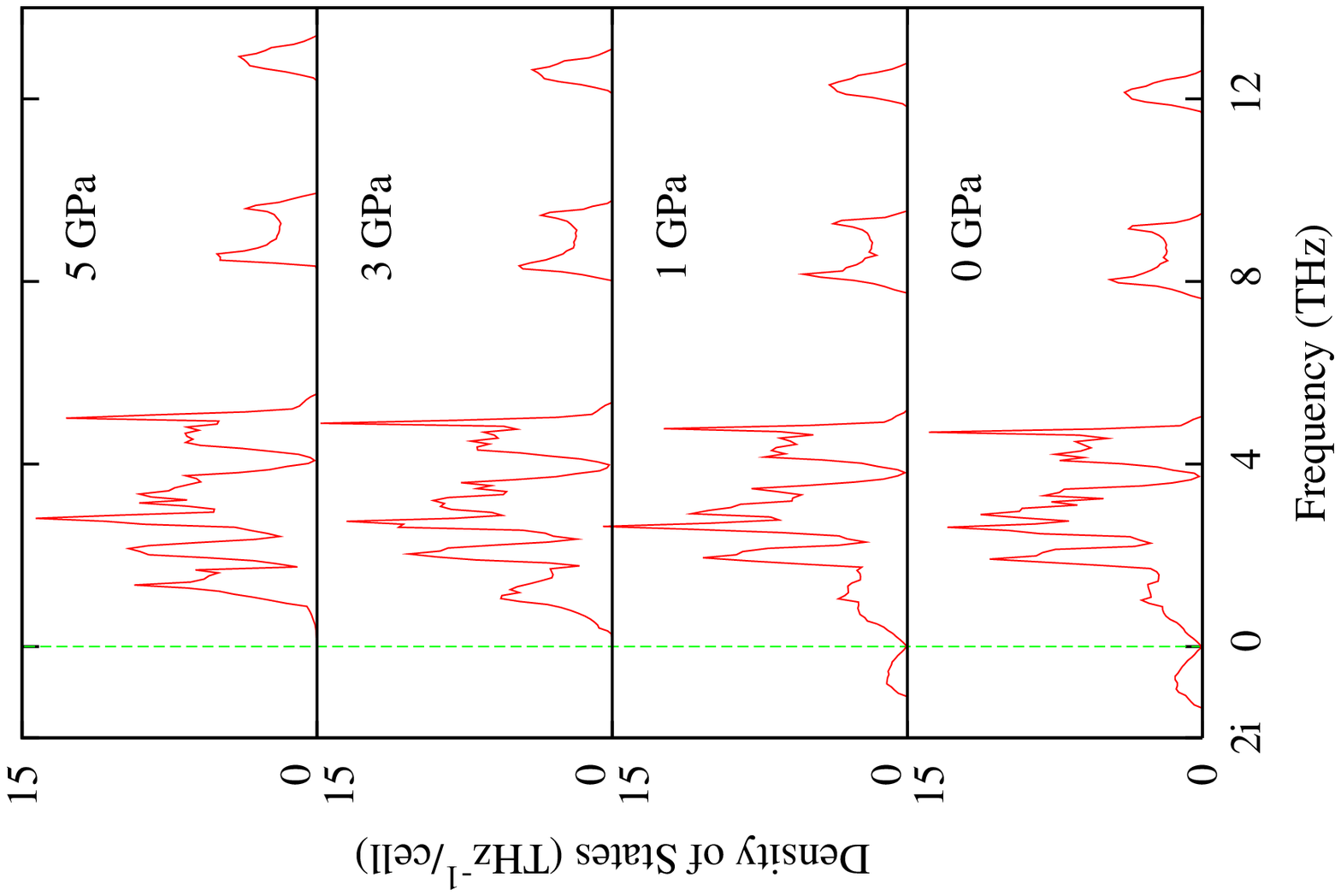}}}}
  \end{minipage}
  \caption{(a)-(b): Phonon band structure of SrPt$_3$P at (a) 0 GPa and (b) 3 GPa. (c): Phonon density of states under different pressures. Negative frequencies indicate the imaginary (soft) phonon modes caused by CDW instability. \label{fig_phonon}}
\end{figure}

\subsection{Phonon Properties and Charge-Density-Wave Instability}
We further investigate the mechanical properties of these materials
by performing phonon calculations using the frozen phonon method.
For SrPt$_3$P, the resulting phonon band structure and DOS exhibit a
clear instability evidenced by the appearance of the imaginary
phonon frequencies. Remarkably, such instability is absent in
CaPt$_3$P. Usually, two scenarios can be responsible for this
instability, namely, the SDW and CDW, respectively. The former
appears frequently in the $3d$-transition-metal pnictide
superconductors. We performed calculations with four possible spin
configurations including/excluding the SOC. All these calculations
eventually converge to the non-magnetic spin configuration. Thus we
exclude the SDW instability in this system.

Noting that the soft phonon modes are located around the $X$ 
($\pi$, $\pi$, 0) and $R$ ($\pi$, $\pi$, $\pi$) points (FIG. \ref{fig_phonon}(a)), we then used a
$\sqrt{2}\times\sqrt{2}\times 1$ supercell with respect to its 
original unit cell to adapt the lattice
distortions due to the possible CDW instability. A less than $0.2\%$
distortion to the atomic coordinates with
$\sqrt{2}\times\sqrt{2}\times 1$ modulation is found to lower the
total energy by $\sim$ 10 meV/cell (or $\sim$ 5 meV/formula). It is
also confirmed that the soft phonon mode disappears if the distorted
supercell is used as the static initial condition to calculate
phonon properties. Therefore, we demonstrated that the soft phonon
modes are in fact due to the CDW instability in SrPt$_3$P. It is
important to emphasize that the energy scale of the predicted CDW
instability is about $\sim$ 1 meV/atom,
which is at least one order smaller than the SDW energy scale in
iron pnictides. This indicates the CDW temperature is roughly around
10 K or even lower in SrPt$_3$P. Further calculations show that the 
weak CDW instability is greatly suppressed by applying an external 
pressure and disappears above 3 GPa.

We next investigate the correlation between the CDW instability and
the crystal structure by showing the evolution of the phonon DOS of
SrPt$_3$P under the external pressure(FIG. \ref{fig_phonon}). At 
zero GPa, a DOS hump below 0 THz is clearly presented due to the
soft phonon modes caused by the CDW instability as shown above. The
first real phonon DOS peak is formed around 2 THz, primarily
contributed by the lowest optical modes. The external pressure
compresses the interatomic distance, causing stiffer bonds, and
therefore all real phonon modes shift to higher energies. For the
soft modes, the DOS hump below 0 THz is reduced at 1 GPa and
eventually disappears at 3 GPa, indicating that the CDW instability
can be quickly suppressed by the external pressure. The {\it
hardened} soft modes contribute to the states below 2 THz, forming
the new first peak below 2 THz.

\begin{figure}[htp]
  \subfigure[$\lambda_{\mathbf{q}}$] {
    \rotatebox{270}{\scalebox{0.5}{\includegraphics{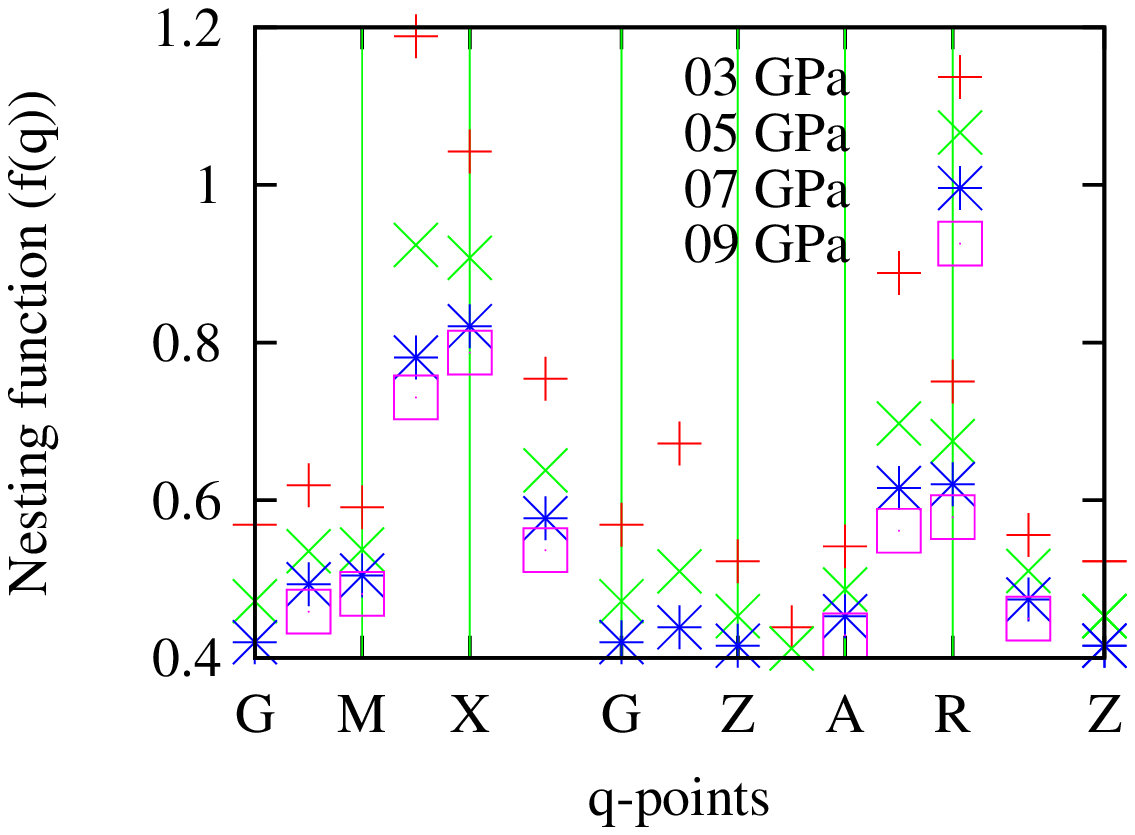}}}}
  \subfigure[$T_c$] {
    \rotatebox{270}{\scalebox{0.5}{\includegraphics{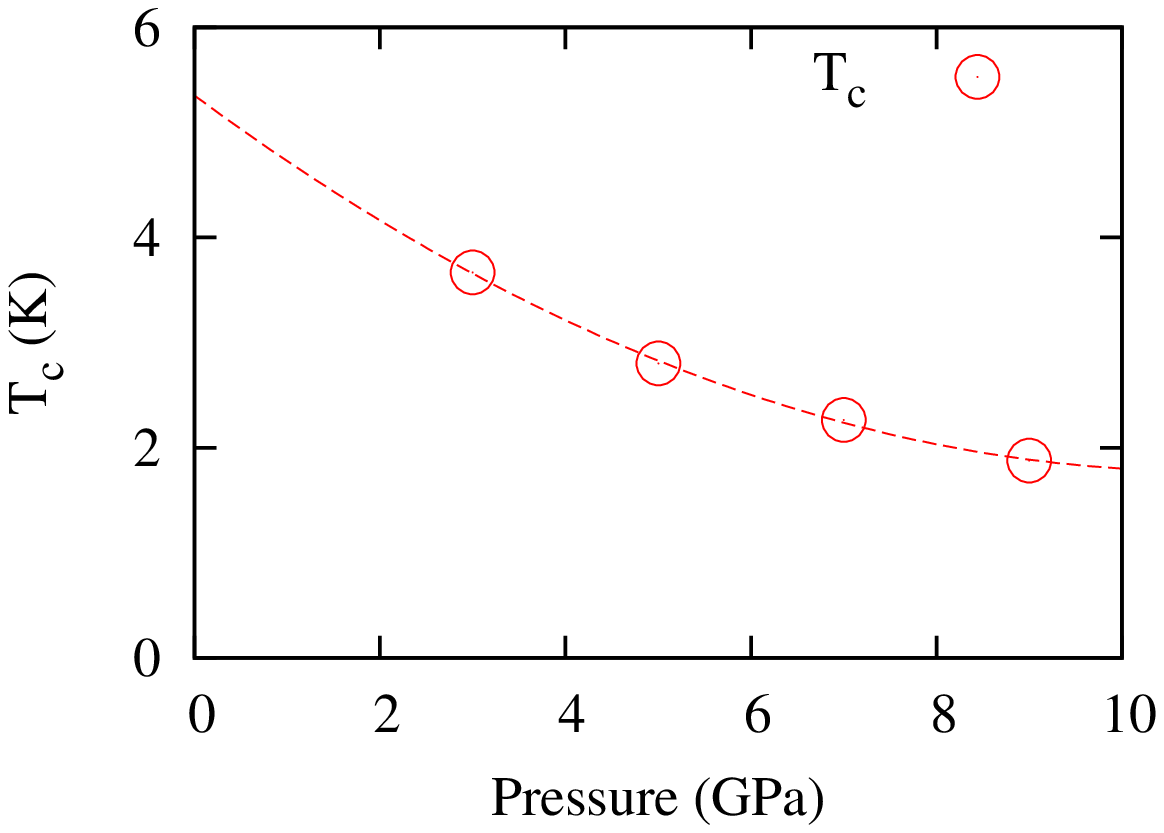}}}}
  \caption{(a) $\mathbf{q}$-dependent electron-phonon coupling constant $\lambda_{\mathbf{q}}$ of SrPt$_3$P under different pressures. (b)SC transition temperature $T_c$ of SrPt$_3$P at different pressures. The circles in (b) indicates the $T_c$ obtained by first-principles electron-phonon coupling calculations and the Allen-Dynes formula; and the dashed line is a quadratic curve fitted to the results. \label{fig_elph}}
\end{figure}

To analyze the formation of the CDW instability, we calculated the
Fermi surface nesting function $f(\mathbf{q})=\frac{2}{N_\mathbf{k}}
\sum_{ \mathbf{k}mn } \delta (\epsilon_{ \mathbf{k+q}n } ) \delta (
\epsilon_{\mathbf{k}m } )$ and the non-interacting electron response
function
$\chi_0(\mathbf{q})=\frac{1}{N_{\mathbf{k}}}\sum_{\mathbf{k}mn}\frac{f_{\mathbf{k+q}n}-f_{\mathbf{k}m}}{\epsilon_{
\mathbf{k+q}n }-\epsilon_{ \mathbf{k}m }}$ for SrPt$_3$P at different pressures, as well as for
CaPt$_3$P at 0 GPa (FIG. \ref{fig_fsnest})\footnote{The Fermi
surfaces, nesting functions, and the non-interacting electron
response functions are calculated by fitting the LDA band structure
using 64 maximally localized wannier orbitals and then interpolate
the Hamiltonian on a $100\times100\times100$ $\Gamma$-centered
K-mesh.}. In all these cases, no significant divergence
appears in both functions. Instead, $f(\mathbf{q})$ shows peaks
not only around $X$, but also around $M$, and $Z$. In the
cases of SrPt$_3$P at 5 GPa and CaPt$_3$P at 0 GPa, the nesting
functions show even stronger peaks around $Z$. Besides,
$\chi_0(\mathbf{q})$ shows broad humps instead of sharp peaks around
$X$, and that feature does not show much variation. Thus the
Fermi surface nesting effect alone cannot be responsible for the CDW
instability. Recently, it was suggested that in the dichalcogenides
the CDW could be a result of strong electron-phonon coupling at
certain $\mathbf{q}$
vectors\cite{weber_prl_107,matteo_prl_106,PhysRevB.73.205102}. If
this is the case here, the $\mathbf{q}$-resolved electron-phonon
coupling constant $\lambda_\mathbf{q}$ of SrPt$_3$P should exhibit
an apparent anomaly around the CDW wave vector at 0 GPa, which would
then be suppressed by external pressure and disappear beyond 3 GPa.
In order to clarify this possibility, we performed first-principles
electron-phonon calculation at different pressures (FIG. \ref{fig_elph}(a)). The resulting wave vector,
however, is around ($\pi$/2, $\pi$, 0), far from the predicted CDW vector
($\pi$, $\pi$, 0). Therefore, the electron-phonon coupling alone cannot
be accounted for the present CDW instability either.

\subsection{Superconducting $T_c$ and Specific Heat Anomaly}
We now discuss the $T_c$ of SrPt$_3$P within the BCS theory. A
direct calculation of $T_c$ is prohibited due to the presence of the
CDW-induced imaginary phonon modes. We thus performed a series of
calculations at the pressures from 3 to 9 GPa, where the CDW
instability is absent. The variation of the BCS $T_c$ under the
external pressure $p$ is obtained by using the Allen-Dynes
formula\cite{allen_dynes}, and then extrapolated to $p=0$ with
least square fitting method assuming a quadratic dependence of
$T_c$ on $p$(FIG. \ref{fig_elph}(b)). Hence the extrapolation represents the $T_c$ without
the CDW instability. The resulting transition temperature is 5.4 K,
much lower than the 8.4 K as observed in experiment. The large 
difference between the calculated and experimental $T_c$ may be an
evidence for CDW-enhanced superconductivity.

In addition to $T_c$, the experiment reveals that the specific heat
coefficient $\gamma$, which is proportional to the electron DOS at
the Fermi level $g(E_F)$, is relatively larger in CaPt$_3$P. Our
calculation shows that in the absence of the CDW, $g(E_F)$ is 4.58 eV$^{-1}$/cell
for SrPt$_3$P, which is slightly larger than the corresponding value
4.53 eV$^{-1}$/cell for CaPt$_3$P. We then performed the DOS calculations using a
$\sqrt{2}\times\sqrt{2}\times1$ super cell that accommodates CDW
distortions. The resulting $g(E_F)=4.31$ eV$^{-1}$/cell, which is 6\% smaller than
the one without CDW. Thus the formation of the CDW indeed leads to a
sizable reduction in $\gamma$, in agreement with the experiment. The
experimental data also reveal other delicacies related to the
specific heat and resistivity in SrPt$_3$P slightly above the $T_c$.
While these features remain questionable,
we note that these features are just within the energy scale $\sim
10$K where the CDW instability starts to play a role.

Several remarks are then in order. First, the CDW instability
predicted for SrPt$_3$P is rather unexpected. This is because of not
only the relatively high $T_c\sim 8.4$K observed in the $5d$
transition metal compounds, but also the very small CDW energy scale
$T^*\sim 10$K compared to the canonical CDW superconductors such as
in a series of transition metal dichalcogenides
TiSe$_2$\cite{wilson_adv_phys_24,morosan_nphys_2_544,PhysRevLett.98.117007}
and NbSe$_2$\cite{PhysRevLett.34.734,Yokoya21122001}, as well as
another platinum pnictide compound
SrPt$_2$As$_2$\cite{JPSJ.79.123710}. So far the minimal difference
between the CDW and SC temperatures in known CDW materials is
observed in NbSe$_2$ with $T_c=7.2$K and $T^*=32$K.  Second, while
the CDW order competes and sometimes co-exists with the SC in most
of known CDW superconductors, the CDW in these cases is {\it
static}, in the sense that the charge redistribution and ionic
modulation are already stabilized and occur far above the SC
instability. In the present case, however, the CDW instability
should be considered as {\it dynamical}, because only a very small
distortion (less than 0.2\%) in atomic coordinates is required to
avoid the imaginary phonon modes and accommodate the lattice
modulation, resulting in a very small characteristic temperature
($\sim 10$K). Perhaps this is the reason why the CDW in SrPt$_{3}$P
has not been reported so far experimentally. It is also possible that 
quantum fluctuations may influence the weak CDW, resulting in quantum 
paraelectric states, similar to the ferroelectric state of SrTiO$_3$. 
Third, in contrast to
the conventional mechanism of the CDW,  neither pure nesting effect
nor the momentum-dependence of the electron-phonon coupling can
account to the CDW formation in SrPt$_3$P independently. This
implies that an appropriate combination of the nestings, their
frustrations, the electron-phonon coupling, or even the states far
away from the Fermi-surface, is at work in stabilizing the CDW in
realistic materials\cite{PhysRevB.77.165135}. We argue that this may
be a characteristic feature of the dynamical CDW.

\section{Conclusion}
In conclusion, we have performed a detailed study on $A$Pt$_3$P
using the first-principles simulations. Our calculations show
that the Fermi surfaces of these materials
consist of multiple sheets. The spin-orbit coupling
effect is negligible in both SrPt$_3$P and CaPt$_3$P but plays an
important role in LaPt$_3$P. We predict that SrPt$_3$P exhibits a
CDW instability which can be quickly suppressed by the external
pressure and is absent in CaPt$_3$P under ambient pressure. The
formation of the CDW is neither directly related to the nesting
effect nor to the electron-phonon coupling alone. Our results
suggest that SrPt$_3$P is a rare material where the SC
instability falls into the fluctuation regime of the dynamical CDW
which in turn leads to the enhancement in $T_c$.

{\it Acknowledgement.} 
The authors would like to thank B. Chen, X.-Y. Feng and N.L. Wang
for valuable discussions. This work was supported in part by the NSF
of China (No. 11274006, 11274084, and 11104051),
the 973 Project of the MOST, and the NSF of Zhejiang
Province (No. LR12A04003 and Z6110033). All the calculations were
performed at the high performance computing center of Hangzhou
Normal University.

{\it Note added.} During the preparation of this manuscript, we
became aware of another work by I. A. Nekrasov and M. V.
Sadovskii\cite{nekrasov_1} on the electronic structure of SrPt$_3$P.
However, they did not consider the spin-orbit coupling effect, the
phonon properties, nor the possible CDW instability in this
material.

\end{document}